\documentclass[aps,twocolumn,floats,floatfix,showpacs,superscriptaddress]{revtex4}
\usepackage{graphicx,amssymb,amsmath}
\usepackage[latin1]{inputenc}
\usepackage{color}

\begin{document}

\title{Forecasting Extreme Events in the Complex Dynamics of a Semiconductor Laser with Feedback}

\author{Meritxell Colet}
\affiliation{Department of Physics and Astronomy, Carleton College, Northfield, 55057 MN, USA}
\author{Andr\'{e}s Aragoneses}
\affiliation{Department of Physics and Astronomy, Carleton College, Northfield, 55057 MN, USA}

\date{\today}

\begin{abstract}
 
Complex systems performing spiking dynamics are widespread in Nature. They cover from earthquakes, to neurons, variable stars, social networks, or stock markets. Understanding and characterizing their dynamics is relevant in order to detect transitions, or to predict unwanted extreme events. Here we study the output intensity of a semiconductor laser with feedback, in a regime where it develops a complex spiking behavior, under an ordinal patterns analysis. We unveil that the complex dynamics presents two competing behaviors that can be distinguished with a thresholding method, and we use temporal correlations to forecast the extreme events, and transitions between dynamics.

\end{abstract}


\maketitle


Nature presents many physical systems where the interplay between a deterministic behavior, stochasticity and time delay leads to a broad variety of complex dynamics \cite{2009_Mitchell,2012_Nat_Phys_Crutchfield,2017_Charbonneau}. These complex systems, constituted by numerous elements interacting non-linearly, present collective emergent phenomena that can not be explained by analyzing its elements individually, but a broader approach is necessary to unveil any hidden structure in its dynamics. Some complex systems manifest their emergent behavior through sequences of extreme oscillations or spiking events. This type of behavior can be found in earthquake activity \cite{1997_Science_Geller,2004_PRL_Corral,2017_SERRA_Kagan}, neuronal dynamics \cite{2005_PRE_Neiman, 2016_Nat_Phys_Sancristobal}, social networks \cite{2014_Sci_Rep_Palchykov}, heartbeat behavior \cite{1993_PRL_Havlin,2012_CBM_parlitz,2012_entropy_Rosso}, optical systems \cite{2013_RMP_Soriano}, stock markets \cite{2012_entropy_Rosso, 2009_PhsA_Zunino}, among others \cite{2015_PRL_Kohar, 2017_PRL_Ginzburgr,2016_Rogue_Waves}.

Understanding and characterizing the different dynamic regimes that a given system can manifest is transcendent to forecast unwanted extreme events, or to distinguish between two competing behaviors that can lead to undesired events  \cite{1997_Science_Geller,2013_PRL_Gauthier, 1996_IJF_Cecen, 2017_EPJST_Masoller}.

Here we study the complex dynamics of the output intensity of a semiconductor laser with optical feedback, in the spiking regime of Low Frequency Fluctuations (LFF). We find that, i) at the onset of the LFF regime, and at the transition from the LFF regime to the coherence collapse regime, the dynamics is characterized by two competing behaviors that can be identified with a thresholding method; and ii) in these transition regimes, temporal correlations in the global spiking dynamics can be used to forecast extreme events, and transitions between types of events.


Semiconductor lasers with optical feedback have shown to manifest a wide range of complex dynamics, from periodicity to high dimensional chaos \cite{2013_RMP_Soriano}. Control and entrainment of these dynamics has practical applications, from encrypted telecommunications \cite{2005_Nature_Ojalvo}, or subwavelength position sensing \cite{2013_OL_Cohen}, to reservoir computing \cite{2013_Nat_Comm_Soriano}. One particular complex dynamics that semiconductor lasers with feedback can exhibit is the Low Frequency Fluctuations \cite{1977_IEEE_Risch}. In this regime, the laser shows a spiking behavior where sudden intensity dropouts happen followed by slow recoveries.

This spiking behavior is consequence of the interplay between nonlinear light-matter interactions, time delay from feedback, and spontaneous emission noise \cite{1980_Lang_Kobayashi}. This behavior takes place for low to moderate optical feedback and around the emission threshold of the laser. As we increase the pump current of the laser the LFF dynamics yields to coherence collapse, where the oscillations are too fast and irregular, and the dropouts cannot be distinguished.

\begin{figure}[tph]
   \centering
     \resizebox{1.0\columnwidth}{!}{\includegraphics{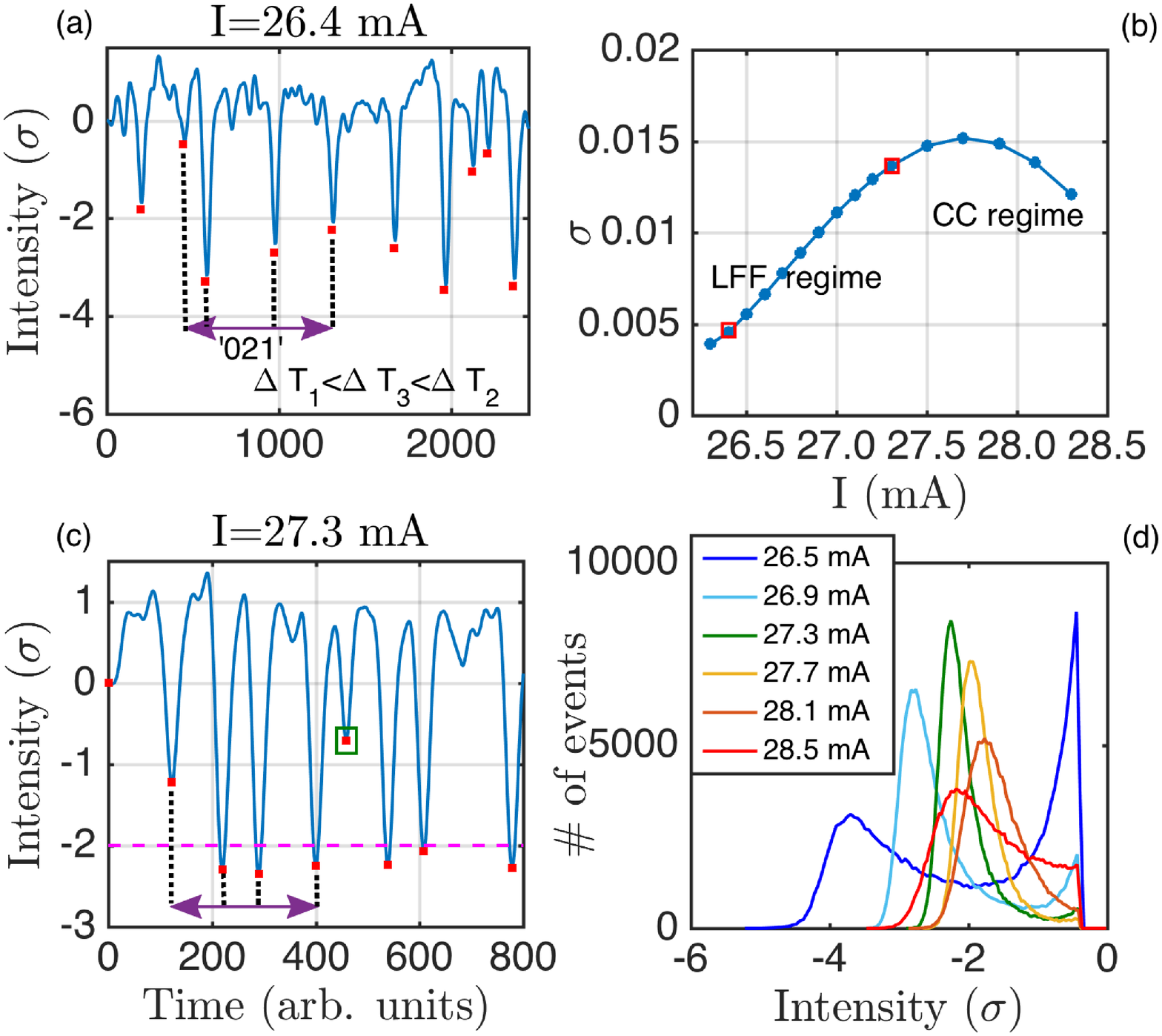}}
    \caption{{\textbf{(a)}} Time series of the output intensity of the laser for $I=26.4~mA$. Red squares indicate the detection of the dips. One word is shown, '021' considering the time intervals of the events (dips). There is a broad distribution in depths of the dips. {\textbf{(c)}} Time series for $I=24.3~mA$. The dips are less spread-out. One word (double arrow) that forecasts the change from deep events to a shallow event (green square) is shown as example (see main text). Threshold is set at $th=-2\sigma$. {\textbf{(b)}} Standard deviation of the time series versus pump current. The uniform increase in $\sigma$ corresponds to the region of LFFs. The change in trend indicates where coherence collapse (CC) begins. Red squares corresponds to the pump currents of the time series of (a) and (c). {\textbf{(d)}}  Histograms of the depths of the dips for various pump currents. As pump current increases the distribution goes from two-mode spread-out to single mode narrow to broad.
     \label{fig_time_series}}
\end{figure}


Figures \ref{fig_time_series}a and  \ref{fig_time_series}c show typical time series of the semiconductor laser with feedback in the LFF regime, for two values of the pump current. Time series have been normalized to have zero mean and unit variance. The dropout events are indicated with red squares. In our study we scan the pump currents in the regime where the LFF dynamics is present (from 26.3 mA to 28.7 mA). We increase the pump current of the laser from the onset of the LFFs, for low pump currents, through well developed LFFs, and to coherence collapse, at higher pump currents. Details of the experimental setup and the laser can be found in \cite{2014_sci_rep_aragoneses}. Figure \ref{fig_time_series}b shows the standard deviation, $\sigma$, of the time series. The LFF regime is characterized by a uniform increase of $\sigma$ with the pump current. Transitions in the dynamics are characterized by a change in the trend of $\sigma$ \cite{2016_sci_rep_Quintero}.

To study the spiking dynamics of the system we use the ordinal patterns analysis introduced by Bandt and Pompe \cite{2002_PRL_BP}. This analysis method transforms a time series of $N$ events into $N-D$ ordinal patterns of dimension $D$, also referred to as words. These words are computed by comparing consecutive inter-event time-intervals, $TI(i) = t(i)-t(i-1)$, where $TI$ refers to time interval, and $t$ is the time where an event occurs. The number of different words depends on the dimension of the words, $D!$. For dimension $D=2$ we have two words: '01' for $TI(i)<TI(i+1)$, and '10' for $TI(i+1)<TI(i)$. For dimension $D=3$ we have six words: '012' for $TI(i)<TI(i+1)<TI(i+2)$, '021' for $TI(i)<TI(i+2)<TI(i+1)$, etc. Figure \ref{fig_time_series}a depicts one word as example.

This method has been shown to be efficient unveiling time correlations in complex time series \cite{2017_pre_kane,2017_entropy_bandt,2013_prl_bandt,2017_PLA_Rosso,2009_EPJB_Masoller}.

A preliminary analysis of the distribution of the events (see Fig. \ref{fig_time_series}d for the histograms computed with the depths of the dips) shows that, as we increase the pump current the distribution of the events changes shape, from a two-mode spread-out distribution to a single-mode narrow distribution, and back to a long tail distribution. This suggests that, for low and high pump currents, the dynamics might be generated by two competing behaviors, one that triggers shallow events and another that triggers deep events. In a previous paper \cite{2013_Sci_Rep_Aragoneses} it was shown that the intensity dropouts of this optical system are triggered by stochastic noise and by an underlying deterministic behavior, showing different statistical behavior.

In order to determine and distinguish the two dynamics we consider a depth threshold to separate them into shallow and deep events (Fig. \ref{fig_time_series}c shows a threshold of $-2\sigma$ as example), and we calculate the probabilities of the words of dimension 6.

\begin{figure}[tph]
   \centering
     \resizebox{1.0\columnwidth}{!}{\includegraphics{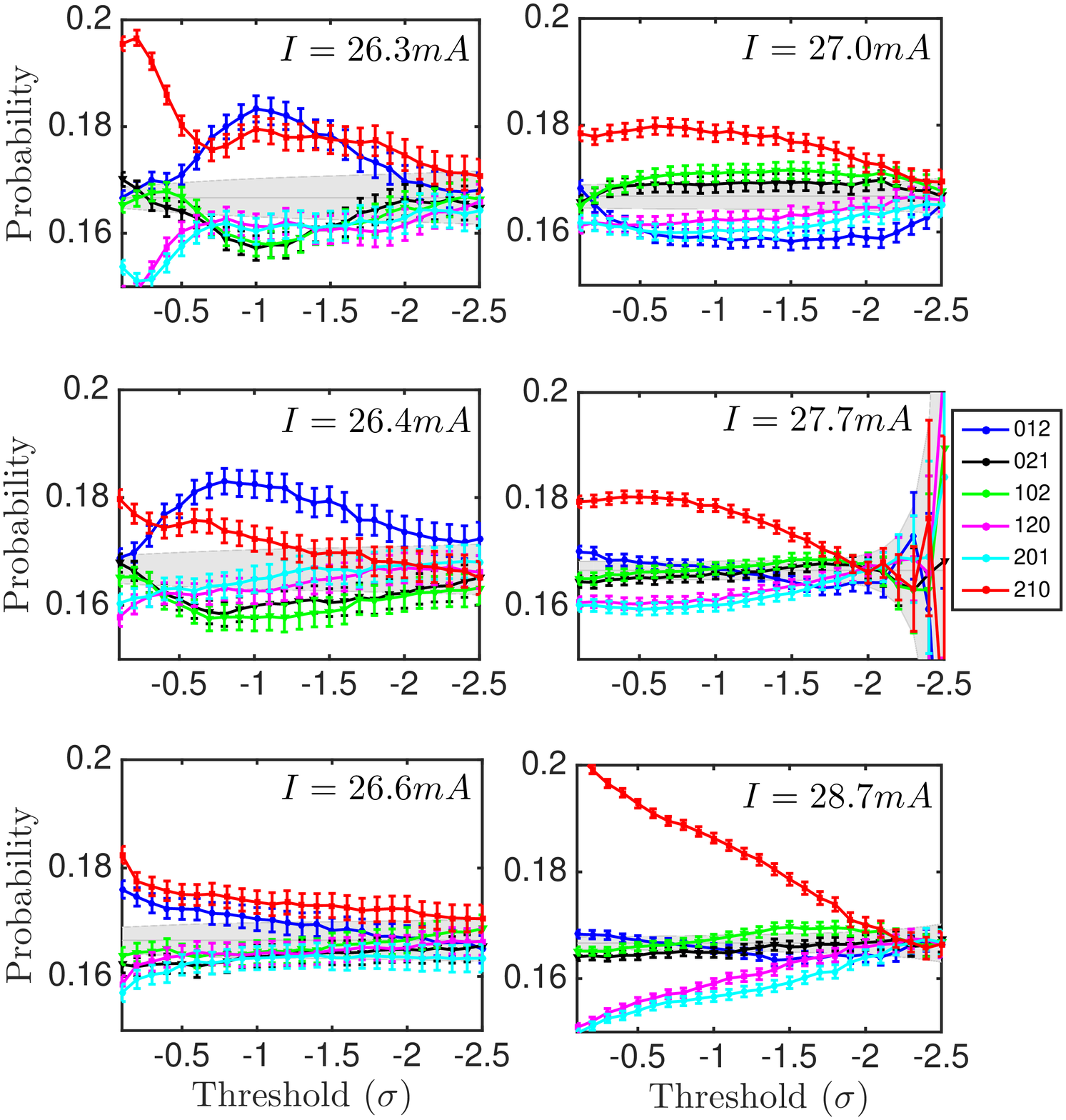}}
    \caption{Words probabilities of dimension 6 versus threshold, computed with the time intervals between events, for different pump currents. Only events deeper than the threshold are considered. The error bars represent the confidence interval computed with a binomial test, corresponding to a confidence level of $95\%$.
     \label{fig_words_th}}
\end{figure}

Figure \ref{fig_words_th} shows the words probabilities versus threshold for different pump currents. The words are computed considering the time intervals between events (dips), and only events below the selected threshold are considered. The gray region corresponds to the probability values consistent with the null hypothesis, that there are no time correlations in the sequence of dropouts and thus, all the words are equally probable ($p \pm 3\sigma_p$, where $p=1/D!$ and $\sigma_p=\sqrt{p(1-p)/N}$, being $N$ the number of words in the sequence).

For low pump currents (onset of the LFFs) the hierarchy of the words show a dependence on the chosen threshold, there is a transition as we lower the detection threshold and exclude shallower events. The temporal correlations between shallow events are different to the ones between deep events, pointing to a dual dynamics. If the events were due to the same underlying dynamics, excluding events would dilute temporal correlations and we would expect the probabilities to drop to the gray region.

For intermediate pump currents (well developed LFFs) the hierarchy does not change with threshold, and the probabilities are closer to the gray region. For high pump currents and shallow thresholds, the probabilities are further away from the gray region than for intermediate pump currents, indicating a more deterministic behavior. In all cases, lower thresholds correspond to a behavior compatible with a stochastic process (probabilities within the null hypothesis region), all temporal correlations are lost.

In many physical systems it is valuable to be able to predict changes in the dynamics, and to forecast extreme events in their dynamics, where the definition of extreme event depends on each particular system \cite{2017_PRL_Ginzburgr, 2016_Rogue_Waves, 2013_PRL_Gauthier,2015_JNCA_Calzarossa, 2017_EPJST_Masoller}. In our optical system, because the global dynamics can be seen as the result of the interaction and competition of two behaviors, it is interesting to be able to forecast changes in the type of events.

In order to forecast the deeper events, but also the change from one type of dynamics to the other (from shallow to deep events and vice versa), we calculate the probabilities of the words that happen right before the transition shallow-to-deep ($X(i)$ is a deep event, $X(i-1)$ is a shallow event, and the word is computed with $[TI(i-3)~TI(i-2)~TI(i-1)]$). We define deep and shallow events as those lower and higher than the threshold, respectively. We also calculate the probabilities of the words that happen right before the transition deep-to-shallow (see Fig. \ref{fig_time_series}c for an example).

\begin{figure}[tph]
   \centering
     \resizebox{1.0\columnwidth}{!}{\includegraphics{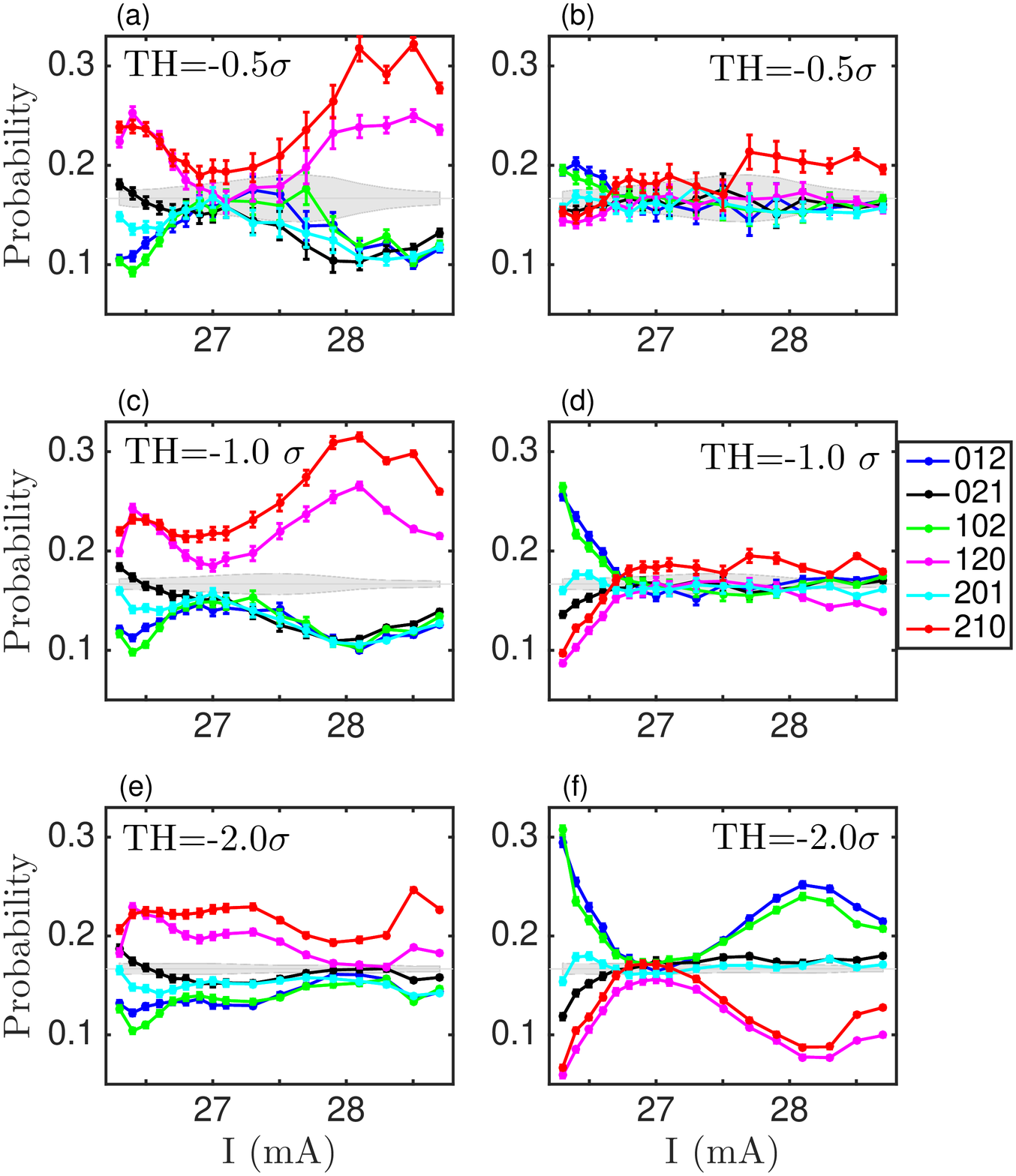}}
    \caption{Words probabilities that forecast the change in dynamics. {\textbf{(a,c,e)}} Forecast from a shallow event to a deep event.  {\textbf{(b,d,f)}} Forecast from a deep event to a shallow event. Three different detection thresholds are shown, $-0.5\sigma$ (a,b), $-1.0\sigma$ (c,d), and $-2.0\sigma$ (e,f).
     \label{fig_predictions}}
\end{figure}

Figure \ref{fig_predictions} shows the probabilities of words of dimension 3 that forecast a change in the dynamics. The left row corresponds to the probabilities of the words that take place before the change shallow-to-deep, i.e., the word that occurs before a deep event, as far as event previous to the deep one is a shallow event. The right row corresponds to the probabilities of the words that take place before the change deep-to-shallow.

Figure \ref{fig_predictions} shows that clear temporal correlations are present before the system goes from one type of dynamics to another. For low and high pump currents, where the dual dynamics is manifest, the system tends to perform some preferred words before it goes from shallow to deep ('120' and '210') and from deep to shallow ('012' and '102'). The forecast depends clearly on the choice of threshold, $th$, as a suitable threshold will separate better the two types of events, while a poor choice of threshold will classify events wrongly. For the shallow-to-deep prediction, $th=-0.5\sigma$ assures that the events above $th$ are shallow events and no deep events are considered as shallow. For the deep-to-shallow prediction, $th=-2.0\sigma$ assures that the events below $th$ are deep events and no shallow events are considered as deep.

In the central range of pump currents, where the LFFs are well developed, and their depths are similar, the forecast is not possible, all six words are equally probable, not showing strong temporal correlations.

\begin{figure}[tph]
   \centering
     \resizebox{1.0\columnwidth}{!}{\includegraphics{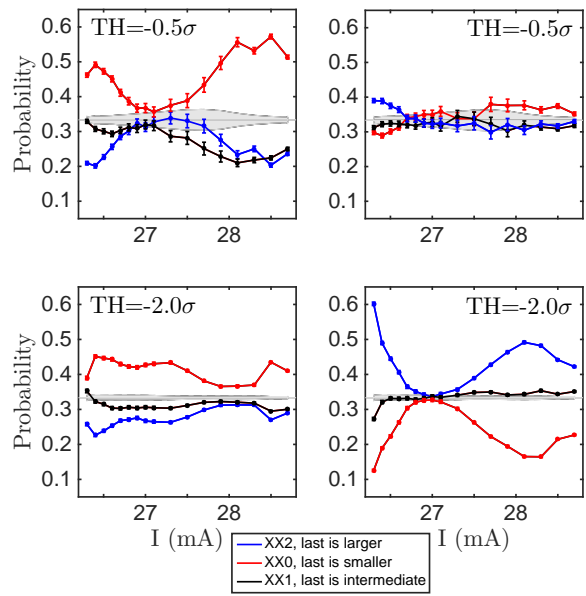}}
    \caption{Probabilities of the forecasting time intervals combinations $XX2=012+102$, $XX1=021+201$, and $XX0=210+120$. These combination highlight the fact that, independently of the two preceding time intervals, the last one tends to be larger or shorter before the transition in the dual dynamics regime. These temporal correlations are lost in the well-developed-LFFs regime.
     \label{fig_predictions_short}}
\end{figure}

Because the preferred words and less preferred words before a change in dynamics are those for which the last time interval is larger or shorter than the preceding two time intervals, we plot the probabilities of the combinations of words that show this global behavior in Fig. \ref{fig_predictions_short} ('XX2=012+102', 'XX0=120+210', 'XX1=201+021'). 

It is in the dual-dynamics regimes where the system allows to forecast changes in the dynamics. There, the temporal correlations present before the deep-to-shallow transition are opposed to those before the shallow-to-deep transition. For some pump currents more than $50\%$ of the transitions occur after the same combination of time intervals ('XX0' for Fig. \ref{fig_predictions_short}a, 'XX2' for Fig. \ref{fig_predictions_short}d), while the probability of transition after the combination 'XX0' for deep-to-shallow transition can be les than $20\%$. 

The decrease in the predictability for the highest pump currents indicates that the system is no longer in the LFF regime and coherence collapse is starting to be the dominant dynamics. This method only allows to predict transitions in a dual dynamics regime.

To summarize, we have used an ordinal patterns analysis to analyze the complex dynamics of the output intensity of a semiconductor laser with feedback. We have uncovered that in the transition regimes to, and from the LFF regime (low and high pump currents), the complex dynamics of the system is the result of the competition of two different behaviors. These two behaviors trigger shallow and deep events, respectively. For the dual dynamics regimes, we have found strong temporal correlations before the dynamics changes from been shallow to deep or vice versa, that allow us to forecast when the following event will be of a different type than the previous one. The temporal correlations that precede a change in the type of events is opposed if it is from shallow to deep or the other way around.

These results may be applied to other complex systems that present dual dynamics in order to distinguish them and predict the occurrence of each one of them. Particularly, because the complex dynamics of semiconductor lasers with feedback have deep similarities with neuronal dynamics, these findings can help understand how biological neurons compute and process information, and path the way to design optical neurons.\\

This work was supported in part by Carleton College Towsley Endowment.


\end{document}